# Análise e modelagem de jogos digitais: relato de uma experiência educacional utilizando metodologias ativas em um grupo multidisciplinar

**David de Oliveira Lemes[1], Ezequiel França dos Santos[1], Eduardo Romanek[1], Adriano Felix Valente[1], Celso Fujimoto[1]**

[1]*Pontifícia Universidade Católica de São Paulo (PUC-SP), São Paulo, Brasil*
*(dolemes@pucsp.br)*

*Resumo:* O ensino tradicional de engenharia de software é focado em competências técnicas. Estratégias ativas, onde os alunos vivenciam conteúdos e manipulam a realidade, são eficazes. O mercado exige novas competências na transformação digital, lidando com a complexidade de modelar negócios e a interconexão entre pessoas, sistemas e tecnologias. A transição para metodologias ativas, como PBL, traz a realidade do mercado para a sala de aula. Este artigo relata a experiência na disciplina, apresentando conceitos e resultados.

*Palavras-chave*: engenharia de software, desenvolvimento de jogos digitais, problem based learning

## INTRODUÇÃO

A disciplina de análise e modelagem de jogos digitais no curso de mestrado profissional em desenvolvimento de jogos digitaIs, da PUC-SP, está voltada para: requisitos de software, mutabilidade de requisitos, modelagem por casos de uso, modelagem de domínio, diagrama de robustez, diagrama de sequência, diagrama de comunicação, diagrama de classe, processo de desenvolvimento, técnicas de entrevista, técnicas de levantamento de dados e prototipação de baixa fidelidade.

Sistemas de software são reconhecidamente importantes ativos estratégicos para diversas organizações. Uma vez que tais sistemas, em especial os sistemas de informação, têm um papel vital no apoio aos processos de negócio das organizações é fundamental que funcionem de acordo com os requisitos estabelecidos. Neste contexto, uma importante tarefa no desenvolvimento de software é a identificação e o entendimento dos requisitos dos negócios que os sistemas apoiarão (WOHLIN & AURUM, 2006).

Jogos digitais são softwares que possuem interfaces de interação com o jogador, mas possuem características distintas de sistemas de softwares comerciais convencionais. A principal diferença é que os jogos são feitos com a finalidade de desafiar o jogador (DE MORAIS & FALCÃO, 2007), gerando diferentes requisitos (narrativa, mecânica, dinâmicas, entre outros). Dessa forma, há a necessidade de adaptar-se às metodologias de desenvolvimento de software para atender a estes novos requisitos, isso demanda diferentes competências e habilidades.

podemos observar as diversas evoluções do mercado de jogos digitais, alinhadas às mudanças comportamentais da sociedade (SANTOS, 2022).

O profissional de desenvolvimento de jogos e engenharia de software no mercado de trabalho atual necessita de novas competências dentro do contexto da transformação digital (FERREIRA et al., 2018), competências chamadas de soft skills, tais como trabalhar em equipe, comunicação, proatividade, criatividade, entre outras (FACQ-MELLET, 2016). Neste novo mercado este profissional enfrenta novos desafios, que vão desde modelar cenários de negócios, o que agrega maior complexidade ao software e à sua construção, a lidar com a interconexão entre pessoas, processos, sistemas e tecnologias (OLIVEIRA, 2018).

No entanto, disciplinas de engenharia de software por muitas vezes não trabalham adequadamente as soft skills, o ensino tradicional dessas disciplinas, em geral, possuem interação e colaboração restrita, com o foco em habilidades técnicas. Ainda neste contexto, estratégias de ensino ativas, como a Problem Based Learning (PBL) têm se mostrado uma boa opção no ensino de engenharia de software, pois propiciam ao estudante a oportunidade de vivenciar os conteúdos apresentados, tirando-os da condição de meros observadores para manipuladores de instrumentos da realidade (LEMOS et al., 2019).

A PBL é uma estratégia educacional centrada no estudante, que o auxilia no desenvolvimento do raciocínio e comunicação, habilidades essenciais para o sucesso em sua vida profissional. Nesta metodologia, o estudante é constantemente estimulado a aprender e a fazer parte do processo da construção do aprendizado (DELISLE, 1997) (SULLIVAN, 1998) (BAYAT & ROHANI, 2012).

Este relato de caso detalha a implementação de metodologias ativas no ensino de engenharia de software, com especial ênfase na análise e modelagem de jogos digitais. As metodologias abordadas, incluindo Aprendizagem Baseada em Problemas (PBL) e estratégias colaborativas, foram adotadas para proporcionar um ambiente educacional mais dinâmico e interativo. O objetivo deste relato é refletir sobre como essas metodologias podem enriquecer o ensino de engenharia de software, alinhando a educação às exigências práticas da indústria de jogos digitais e melhorando as competências dos alunos.

## MATERIAIS E MÉTODOS

### CONTEXTO DA EXPERIÊNCIA DA DISCIPLINA E CONCEITUAÇÕES

O corpo discente desta disciplina do mestrado, como veremos com mais detalhes nas próximas seções, era um grupo heterogêneo, com estudantes oriundos de graduação em tecnologia da informação, estudantes com formação em linhas de design e estudantes com uma formação específica em desenvolvimento de jogos digitais. Isso ajudou a evidenciar as diferenças e similaridades entre uma empresa de software tradicional e uma empresa de jogos. Para tornar estes aspectos mais explícitos, os encontros foram feitos com base em estudos de caso, criando a oportunidade de vivenciar as dificuldades sociais do desenvolvimento, como a comunicação, sem deixar de trabalhar com os aspectos técnicos de modelagem.

Para abranger as diferenças entre um software tradicional e um jogo digital, durante o processo de modelagem foi utilizado o modelo MDA (de Mecânica, Dinâmica e Estética (*Mechanics, Dynamics and Aesthetics*). Cada estudo de caso continha falhas, para que os estudantes questionassem e construíssem uma visão crítica sobre as técnicas de engenharia de *software*, não existindo uma única solução e lembrando sempre que não existe uma "bala de prata" (BROOKS, 1987).

### O ESTUDO DE CASO

Uma empresa fictícia de desenvolvimento de jogos digitais e desenvolvimento de conteúdos digitais interativos passará os próximos seis meses trabalhando com projetos relacionados à disciplina de Física. Os estudantes trabalharam com foco em uma equipe específica da empresa, que era a responsável por criar um jogo digital sobre as Leis de Kepler.

Durante os encontros, novos personagens da equipe eram revelados, com diferente experiência e formação, levando propositalmente às falhas mencionadas na seção anterior, no intuito que os estudantes refletissem e questionassem soluções.

### O FRAMEWORK MDA

O Framework MDA (de Mecânica, Dinâmica e Estética (*Mechanics, Dynamics and Aesthetics*), tem sido amplamente utilizado no processo de desenvolvimento e design de aplicações de jogos. Esse framework corresponde aos principais elementos encontrados nos jogos. No modelo MDA os autores apresentaram uma forma diferente de perceber, analisar e projetar jogos digitais, considerando-os como produtos consumíveis, porém com uma expectativa de experiência de uso imprevisível (DORMANS, 2012). O MDA, por sua flexibilidade, deveria ser "tem sido também utilizado em pesquisa e desenvolvimento (ZAFFARI & BATTAIOLA, 2014), assim como em contextos educacionais (SOUZA et al., 2018).. Durante a disciplina o MDA foi utilizado integrado aos processos de levantamento de requisitos, assim como suporte no desenvolvimento do GDD.

### O GDD

O GDD *Game Design Document* é utilizado para documentação no processo de desenvolvimento de jogos, o que pode incluir a etapa de requisitos e diagramas elaborados. O GDD possui o intuito de facilitar o desenvolvimento do jogo, e não exige *templates* (BREVES et al., 2021) em sua elaboração, possibilitando ser adaptado. Durante a disciplina, a coleção de artefatos compuseram o GDD de cada grupo.

### MÉTODO DE TRABALHO

A disciplina foi ministrada utilizando a estratégia educacional PBL, em encontros semanais de três horas com estudo dirigido e leituras entre as aulas.
Além de trabalhar a partir de problemas apresentados nos estudos de casos, os estudantes eram constantemente estimulados a levantarem novos e adicionais problemas e visões, além de também trabalharem sempre como agentes ativos na solução e proposta de ferramentas.

### GRUPOS MULTIDISCIPLINARES

Em um primeiro contexto, temos uma equipe de desenvolvedores de *software* tradicional, que geralmente é composta inteiramente por profissionais da área de Tecnologia da Informação. Uma vantagem dessa estrutura é que todos possuem uma linguagem comum e um certo entendimento do produto em que estão trabalhando. Em um segundo contexto, temos uma equipe de desenvolvedores das áreas de transformação digital, pesquisa e inovação e desenvolvimento de games, que não têm essa praticidade, por estarem em um contexto multidisciplinar, sendo necessário encontrar uma linguagem que alcance a todos os membros de modo que estes possam trabalhar em conjunto e confiar na capacidade de todos para executarem suas tarefas, para que, no final, possam unir todos os elementos em um único produto (BREVES et al., 2021).

O corpo discente se enquadra no segundo contexto, com estudantes oriundos de graduação em tecnologia da informação, estudantes com formação em linhas de design e estudantes com uma formação interdisciplinar em jogos digitais. Essa heterogeneidade de formações trouxe ao grupo grande troca de experiências, acompanhados de desafios. Por este motivo, e para garantir a heterogeneidade do grupo, durante as aulas, aqui chamadas de encontros, foram definidos dois papeis principais e sete passos a serem seguidos.

PAPEIS DOS ESTUDANTES

Dois papeis principais, que deveriam ser ocupados por alunos diferentes a cada encontro:

**Papel 1**: Coordenador da Sessão do Encontro :
1. Coordenar o processo de trabalho do grupo, incluindo o controle do tempo
2. Introduzir o problema para discussão
3. Facilitar a participação de todos os membros, certificando-se que os alunos exerçam o mesmo grau de participação
4. Contribuir com o grupo e facilitar a discussão
5. Elaborar e estimular a discussão
6. Observar o tempo e direcionar a discussão
7. Certificar-se do cumprimento de todos os passos do PBL

**Papel 2**: Relator na Sessão do Encontro :
1. Prestar atenção na evolução das ideias do grupo
2. Anotar as ideias surgidas no grupo, mesmo que inicialmente pareçam irrealizáveis - princípio da metodologia tempestade de ideias
3. Organizar as ideias, sintetizando-as em conceitos
4. Conferir suas anotações com os outros membros do grupo
5. Contribuir com o grupo e facilitar a discussão
6. Resumir os objetivos de aprendizagem decididos pelo grupo ao final de cada sessão tutorial

PASSOS DOS ENCONTROS NA ESTRATÉGIA PBL

Sete passos seguidos em cada encontro.

Passo 1: Esclarecimento de termos e expressões no texto do problema - Todos os termos desconhecidos e aqueles que permanecerem sem definição satisfatória ao serem discutidos no grupo serão objeto de estudo, devendo ser esclarecidos pelos estudantes na próxima sessão.

Passo 2: Definição do problema
Compreender o foco do problema, sintetizando seus principais aspectos, sua essência.

Passo 3: Análise do problema (tempestade de ideias)
A discussão deverá ser encaminhada para delimitação das questões, procurando-se explicações baseadas nos conhecimentos prévios que os próprios estudantes trazem consigo.

Passo 4: Sistematização das explicações com a proposição de hipóteses - Na sequência, os estudantes proporão justificativas plausíveis para explicar as questões previamente levantadas pelo problema como um todo.

Passo 5: Proposição dos objetivos de aprendizagem - Os estudantes deverão propor quais os objetivos de aprendizagem que validem as hipóteses construídas, ou seja, o que eles devem aprender para resolver o problema.

Passo 6: Estudo individual - O estudante deve identificar as atividades necessárias para a resolução do problema, atingir os objetivos de aprendizagem e proceder à sua aprendizagem. Individualmente, entre uma sessão de tutoria e outra, os estudantes deverão estudar os termos desconhecidos do texto, responder aos objetivos levantados na sessão de abertura e demais assuntos que faltaram para completar a análise do problema. Sistematizando seus estudos da forma que julgar mais apropriada.

Passo 7: Fechamento do problema - Após o estudo, ocorre o segundo encontro tutorial. Os estudantes deverão estar aptos a contribuir com a resolução do problema. Devem confrontar os resultados de seus estudos com a produção dos outros colegas e refletir sobre suas próprias hipóteses formuladas durante a autoaprendizado. Com base nas ideias sistematizadas na sessão anterior e na atual, devem surgir as soluções para o problema proposto.

## RESULTADOS E DISCUSSÃO

ARTEFATOS E ENTREGAS

Durante o semestre, cada grupo trabalhou em seu projeto de maneira incremental adentrando os conceitos de modelagem de software alinhado às perspectivas do MDA. Para tanto, um documento designado como GDD (*Game Design Document*) foi sendo elaborado semana a semana através de artefatos construídos a cada encontro como veremos a seguir.

DIAGRAMA DE REQUISITOS FUNCIONAIS E NÃO FUNCIONAIS

Diferente de um software incremental, que é desenvolvido em etapas que podem ou não ser distribuídas como versões, em um jogo de maneira geral a distribuição só é feita ao término do projeto. A engenharia de requisitos é um ramo da engenharia de software que se preocupa com os objetivos, funções e restrições que os sistemas de software devem atender (Sommerville, 2011) (CMMI, 2010).

**Objeto Educacional Interativo: Lei de Kepler**

| Requisitos funcionais | Requisitos não-funcionais |
|---|---|
| Simular os princípios da lei de Kepler | Login do Aluno |
| Um objeto educacional para cada uma das três leis | Tecnologia Baseada em Web |
| | Modelos 3D que podem ser exportados |
| | Integração com Google Earth |

Fig.1: Exemplo de quadro de requisitos desenvolvido por estudante.

ELUCIDAÇÃO DE REQUISITOS E ENTREVISTAS

Após o levantamento dos requisitos vistos na seção anterior, a elucidação destes foi trabalhada através de entrevistas. Além disso, dentro do estudo de caso, a própria equipe fictícia tinha problemas de processo, liderança e comunicação, que também foram debatidos entre os estudantes.

Após a entrevista um protótipo de baixa fidelidade foi desenvolvido e apresentado ao cliente a fim de elucidar questões.

ENTREVISTAS

Nesta etapa, cada grupo deveria criar um formulário que os ajudasse a entender melhor o problema proposto - o ensino das Leis de Kepler - assi[1]m como as necessidades e expectativas do cliente em relação ao produto final.

Em seguida a entrevista foi simulada com os integrantes do outro grupo atuando como os clientes e vice versa.

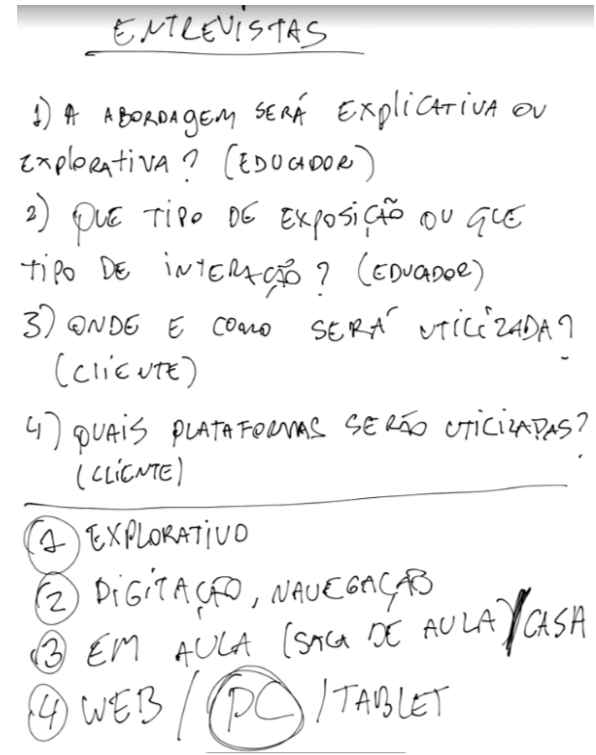

Fig.2: Exemplo de questionário de entrevista desenvolvido por estudante.

PROTÓTIPO DE BAIXA FIDELIDADE

Durante o encontro os estudantes levantaram duas ferramentas de prototipagem para jogos que permitem testar e iterar em tempo real, sendo uma delas o Machinations[1] e a outra o Twine[2]. A primeira opção é ideal para jogos mecânicos onde existem dinâmicas com recursos (DORMANS, 2012) enquanto a segunda opção é mais adequada para jogos de narrativas interativa não lineares (ENGSTRÖM & ERLANDSSON, 2018). Todos optaram pelo Machinations por ser adequado à proposta de jogo que estava sendo elaborada, com narrativas simples e lineares e níveis variados de recursos e processos. Outra vantagem que ambas as ferramentas oferecem é de trabalhar em modo colaborativo online de modo que todos possam contribuir remotamente e avaliar o progresso.

---

[1] https://machinations.io
[2] https://twinery.org

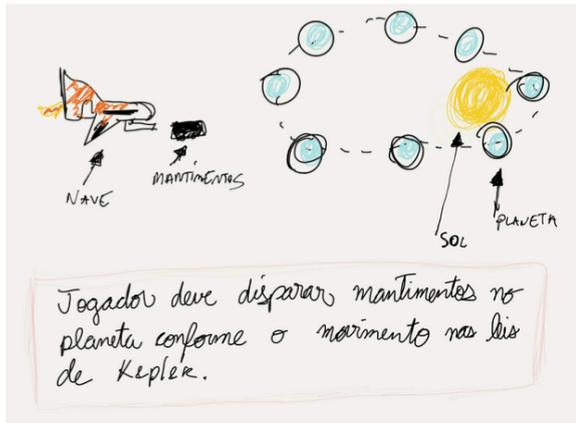

Fig.3: Exemplo de protótipo de baixa fidelidade desenvolvido por estudantes.

## CONFLITOS E COMUNICAÇÃO

Neste encontro foram levantados, discutidos entre os estudantes e posteriormente trabalhados no estudo individual todos os possíveis conflitos entre a proposta e os resultados apresentados até então. Foi realizada uma revisão geral em cada tópico abordado em cada encontro anterior implementando os ajustes necessários.

> Ainda na metodologia, durante o processo de levantamento de requisitos, ponte entre a equipe de desenvolvimento e os *stakeholders*, o que requer aptidões interpessoais para lidar com as possíveis situações de conflito em suas interações, como tomada de decisão, negociação e competências conversacionais, nem sempre, essas habilidades são desenvolvidas no processo de formação do profissional no meio acadêmico. Parte do onboarding da empresa poderia incluir palestras e cursos de gestão de projetos, diversidade e comunicação não violenta.

Fig.4: Exemplo de conclusão levantada por estudante após debate sobre comunicação e liderança.

## GAME DESIGN

Esta foi uma das primeiras etapas que não seriam contempladas em sua totalidade em um curso de engenharia de software tradicional, pois nos softwares convencionais os requisitos usualmente são projetados para facilitar tarefas do usuário, enquanto nos jogos digitais as regras existem para desafiar o jogador (BREVES 2021), (LEMES, 2009).

Para tornar esse processo unificado o modelo MDA (Mechanics, Dynamics and Aesthetics serve para garantir que as mecânicas propostas, com foco em funcionalidade, servirão para gerar dinâmicas de *gameplay* compatíveis com a estética desejada, levando em consideração o público-alvo. Nos estágios iniciais de planejamento, o protótipo de baixa fidelidade tem um papel crucial para o design do jogo. Nos exemplos citados, a ferramenta Machinations facilita uma linguagem em comum para o programador e o game designer, sendo possível explorar conceitos e balancear a dinâmica, enquanto a ferramenta Twine conecta o game designer as disciplinas de narrativa e arte, garantindo uma visão unificada em uma equipe multidisciplinar.

## DIAGRAMA DE CONTEXTO

Em posse do Game Design Document a próxima etapa foi de elaboração do diagrama de contexto.

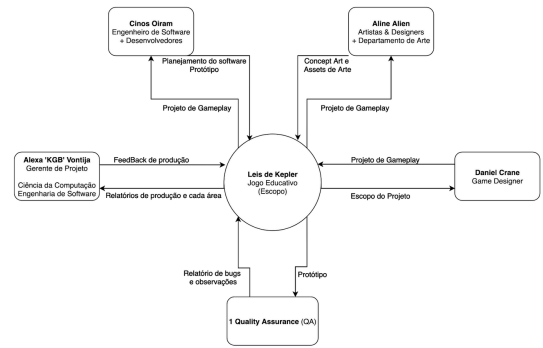

Fig.5: Exemplo de diagrama de contexto desenvolvido por estudantes

## DIAGRAMA DE CASOS DE USO

Uma das ferramentas de modelagem de software que pode trazer grandes ganhos na fase de projeto de jogos digitais é o diagrama de casos de uso. Apesar de não comprovada relação entre aprendizado de diagrama de casos de uso e desenvolvimento de jogos (SILVA, J. C. et al., 2012), a relação entre os dois mostrou-se engajadora para elucidação destes requisitos.

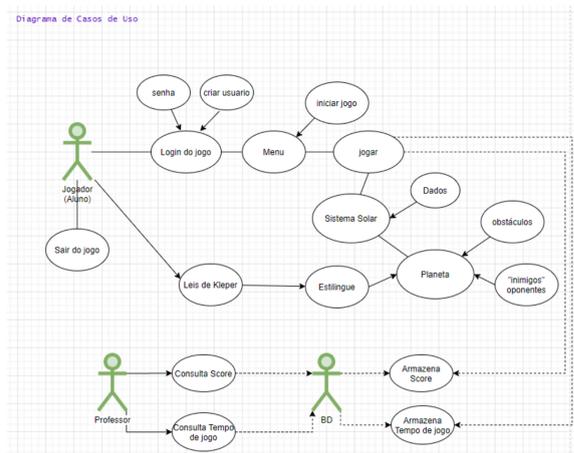

Fig.6: Exemplo de diagrama de caso de uso desenvolvido por estudantes

DIAGRAMA DE ATIVIDADES

Dentre as ferramentas de modelagem, o diagrama de atividades é uma das mais utilizadas para modelagem e análise de aplicações interativas, especialmente quando há a necessidade de integrar os métodos e técnicas propostos pelas áreas de interação humano computador e engenharia de software (SILVA, W et al. ,2020).

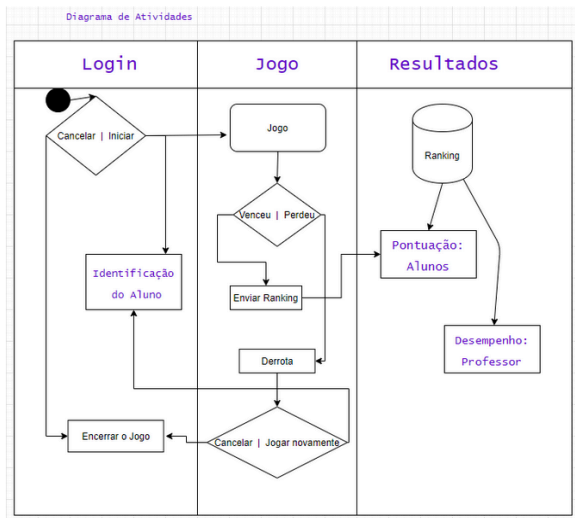

Fig.7: Exemplo de diagrama de atividades desenvolvido por estudantes

DIAGRAMA DE MÁQUINA DE ESTADO

Outra ferramenta de modelagem de software é o diagrama de casos de uso. Composto geralmente por estado, evento, ação, guarda e transição. Este diagrama é de essencial construção e interpretação para contenção e correção de falhas em projetos de software (SILVA, W et al. , 2023).

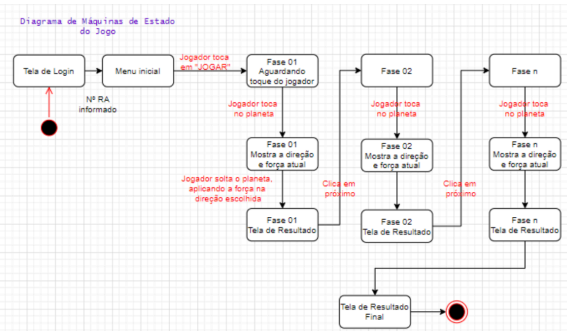

Fig.8: Exemplo de diagrama de máquina de estados desenvolvido por estudantes.

DIAGRAMA DE CLASSES

O diagrama mais utilizado dentro do contexto da UML (*Unified Modeling Language*), pois o mesmo é uma representação da estrutura e relações das classes do software, que servem de modelo para objetos.

Dentro do contexto de educacional, o desenvolvimento de jogos é um dos menos agressivos e práticos para o ensino de modelagem e elaboração de um diagrama de classes, mesmo em cursos tradicionais de engenharia de software (SILVEIRA & SILVA, 2006).

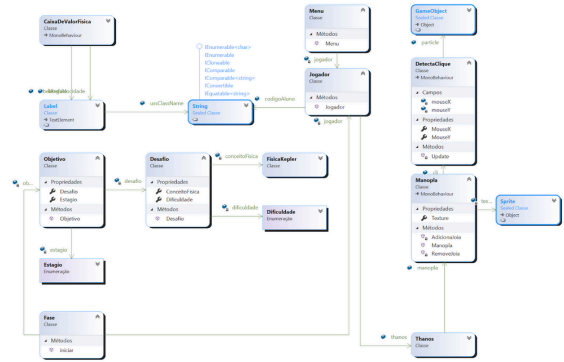

Fig.9: Exemplo de diagrama de classes desenvolvido por estudantes.

RESULTADOS FINAIS OBTIDOS E DISCUSSÃO

O uso de metodologias ativas tem apresentado excelentes resultados quando aplicadas ao ensino de engenharia de software, os estudantes acabam por atingir experiências que só poderiam ser vivenciadas ao ingressar no mercado de trabalho (CAVALCANTI et al., 2008).

Durante a disciplina apresentaram-se os conceitos abordados, o estudo de caso trabalhado, o método de trabalho com papeis e passos dos encontros. Além disso, alguns exemplos dos resultados obtidos como diagramas, protótipos e análises que compuseram o GDD (*Game Design Document*) construído por cada grupo através dos encontros e uma discussão acerca das competências desenvolvidas ao final da disciplina.

As metodologias possibilitam o protagonismo do estudante, com isto, um dos resultados do trabalho foi o levantamento de uma hipótese para avaliação de jogos através de métodos de engenharia aliados aos conceitos de game design.

No decorrer dos encontros um dos tópicos estudados foi a questão do level design. Para Byrne (2005, p.6) o termo é originário do RPG (Role-playing Game e antigas máquinas de Arcade sendo definido como o mapa, missão ou nível composto de eventos com incrementos de dificuldade que suceder-se-ão no jogo durante a partida (BYRNE, 2006).

Nessa etapa, são constituídos os elementos da metodologia MDA e o game designer pode valer-se do método na construção desse mundo, nesse quesito o desenvolvimento dos jogos digitais tem se pautado em novas mecânicas, jogabilidade e narrativas que torna a experiência do usuário mais imersiva, interativa e com uma melhor assimilação das regras estabelecidas na composição do jogo.

Para Braga (2018), esses elementos "[...] espaço, seus limites e as mecânicas que são executadas no seu interior dão forma ao Level do jogo". E ainda, o jogador realiza essa ação dentro do jogo com o intuito de atingir um resultado específico e desejado (LEMES 2018). Tratando-se da construção do level design, temos os artigos de Dan Taylor[2], onde ele apresenta os dez princípios para composição de um bom level design.

Conforme abordado no processo de modelagem MDA, temos as perspectivas do ponto de vista do designer e do ponto de vista do jogador, e se faz necessário analisar as perspectivas desses dois elementos para o melhor desenvolvimento do jogo, nesse caso o level design (HUNICKE, 2004).

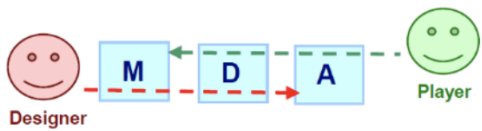

Fig.10: Designer e jogador diferentes perspectivas.
Fonte: (Hunicke, 2004)

Nesse contexto, podemos acrescentar uma nova perspectiva de um terceiro elemento, o ponto de vista do departamento de Garantia de Qualidade (Quality Assurance – QA). Segundo Chandler (2012), a principal responsabilidade do departamento de QA é criar o plano de testes para o jogo e validação do mesmo. A função primordial do QA é executar o playtest do jogo e dar feedback aos desenvolvedores e verificar possíveis bugs durante o processo de análise (MAXWELL, 2012).

O QA consegue avaliar e apontar as necessidades relevantes para melhoria no desenvolvimento do projeto de uma forma técnica concomitantemente com o olhar subjetivo de um jogador, e nisso, temos o refinamento do sistema antes da produção do jogo.

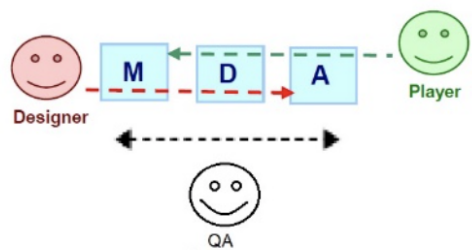

Fig.11: Perspectiva do QA Fonte: Modificado (HUNICKE, 2004) (ROMANEK 2021)

Diante ao exposto, a proposição é a elaboração de uma matriz ponderada como ferramenta de apoio na análise de cada level design, com a finalidade de se ter um parâmetro norteador na melhoria contínua no desenvolvimento do jogo.

Para elaboração dessa matriz, utilizamos o programa em Excel, na qual de forma estruturada quantificamos as variáveis qualitativas de multicritérios do jogo. A concepção da planilha foi realizada a saber: Na primeira matriz "desenvolvimento do jogo", elencamos nas linhas os elementos dos jogos, divididos em três áreas Mecânicas, Dinâmicas, Componentes.

As colunas foram dispostas em:

**Peso**
Através do sistema de pesos definido pelos pares a partir do critério de importância, os valores são 0, 1, 3 ou 5 a ser atribuído a cada um dos itens da respectiva linha.

**Fase**
São as fases do jogo propriamente ditas, onde serão atribuídas as notas em escala numérica de 1 a 7.

**Pontuação máxima**
São os valores de pontuação máxima para aquele item a um determinado peso e nota.

Para a segunda matriz "princípios bom level design" foram elencados nas linhas os dez princípios de um bom level design e em cada coluna foram dispostos os itens (Peso, Fase, Pontuação máxima) da mesma forma supracitada. Como forma de contextualizar o processo, distribuímos os pesos aos itens elencados nos elementos dos jogos, Mecânicas, Dinâmicas, Componentes matriz da Fig.12 e solicitamos ao time de QA atribuir nota mínima 1 até a nota máxima 7 para esses itens. E na matriz Fig. 14, com base nos princípios do bom level design, solicitamos o mesmo procedimento.

Fig.12.: Matriz desenvolvimento jogo (MDC) (Romanek 2021)

Fig.14: Matriz Princípios bom *level* design (PGLD) (Romanek 2021)

Fig.15: Escala numérica Valores de peso dos critérios e nota de avaliação (Romanek 2021)

Com as notas atribuídas conseguimos classificar e visualmente verificar cada quesito das matrizes para cada fase. Com os critérios avaliados, foi gerado um gráfico e determinamos em qual faixa aquele level se encontra, categorizado entre zona baixa, zona intermediária e zona ótima. Assim, pode-se visualizar e auxiliar na tomada de decisão os pontos a serem melhorados. Outra forma de utilização desta ferramenta, é a tomada de decisão para a escolha entre protótipos de alta fidelidade.

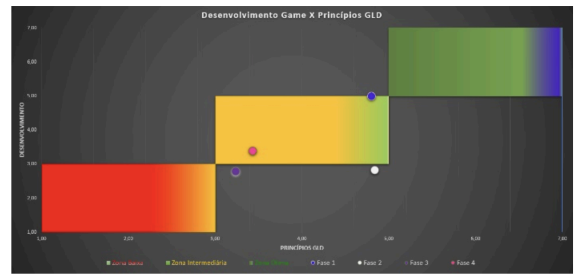

Fig.16: Desenvolvimento Game x Princípios GLD (Romanek 2021)

## CONCLUSÃO

No ensino tradicional de engenharia de software, em geral, existe uma colaboração restrita e com o foco em competências técnicas, os resultados obtidos não seriam possíveis, dada a colaboração unilateral e restrita da relação aluno-professor.

Com uma abordagem construída com metodologias ativas, todo o conteúdo foi contemplado e, dada a construção da turma multidisciplinar e observação dos resultados obtidos, concluímos o potencial de se alcançar além da multidisciplinaridade, a transdisciplinaridade, no momento em que vemos elementos de engenharia de qualidade trabalharem em composição com engenharia de software e game design.

Ainda neste quesito, o PBL (Problem Based Learning se mostra a ferramenta ideal para a preparação dos alunos ao mercado, treinando competências intrínsecas ao ensino puramente das ferramentas de modelagem de software.

Com o emprego das metodologias ativas consegue-se um maior engajamento dos alunos com uma melhor experiência para o desenvolvimento e capacitação profissional.

Dessa forma, temos um ensino-aprendizagem que atende às demandas do mercado, promovendo a aprendizagem de forma colaborativa, reflexiva, por meio de descobertas e soluções.

## REFERÊNCIAS